\documentclass{article}
\usepackage{smc}
\usepackage{times}
\usepackage{ifpdf}
\usepackage[english]{babel}
\usepackage{cite}
\usepackage{xcolor}
\usepackage{soul}
\usepackage{microtype}
\usepackage{paralist}
\usepackage{units}
\usepackage{graphicx}

\def\papertitle{Scream Detection in Heavy Metal Music}
\def\firstauthor{Vedant Kalbag}
\def\secondauthor{Alexander Lerch}
\def\thirdauthor{~}

\newif\ifpdf
\ifx\pdfoutput\relax
\else
   \ifcase\pdfoutput
      \pdffalse
   \else
      \pdftrue
\fi

\ifpdf 
  \usepackage[pdftex,
    pdftitle={\papertitle},
    pdfauthor={\firstauthor, \secondauthor, \thirdauthor},
    bookmarksnumbered, 
    pdfstartview=XYZ 
   ]{hyperref}

  \graphicspath{{./figures/}}
  \DeclareGraphicsExtensions{.pdf,.jpeg,.png,.svg}

  \usepackage[figure,table]{hypcap}

\else 
  \usepackage[dvips,
    bookmarksnumbered, 
    pdfstartview=XYZ 
  ]{hyperref}  

  \usepackage[dvips]{epsfig,graphicx}
  \graphicspath{{./figures/}}
  \DeclareGraphicsExtensions{.eps}

  \usepackage[figure,table]{hypcap}
\fi

\hypersetup{
    colorlinks,%
    citecolor=black,%
    filecolor=black,%
    linkcolor=black,%
    urlcolor=black
}

\title{\papertitle}

\twoauthors
  {\firstauthor} {Music Informatics Group 
  \\Georgia Institute of Technology, USA \\ %
    {\tt \href{vedant.kalbag@gatech.edu}{vedant.kalbag@gatech.edu}}}
  {\secondauthor} {Music Informatics Group 
  \\Georgia Institute of Technology, USA \\ %
    {\tt \href{mailto:alexander.lerch@gatech.edu}{alexander.lerch@gatech.edu}}}

\begin{document}
\capstartfalse
\maketitle
\capstarttrue
\begin{abstract}

Harsh vocal effects such as screams or growls are far more common in heavy metal vocals than the traditionally sung vocal.
This paper explores the problem of detection and classification of extreme vocal techniques in heavy metal music, specifically the identification of different scream techniques. We investigate the suitability of various feature representations, including cepstral, spectral, and temporal features as input representations for classification. The main contributions of this work are
\begin{inparaenum}[(i)]
    \item   a manually annotated dataset comprised of over 280 minutes of heavy metal songs of various genres with a statistical analysis of occurrences of different extreme vocal techniques in heavy metal music, and
    \item   a systematic study of different input feature representations for the classification of heavy metal vocals.
\end{inparaenum}

\end{abstract}
\section{Introduction}\label{sec:introduction}
Vocals in heavy metal music can be very different to those in other styles. Heavy metal vocalists use a variety of techniques, colloquially known as screams or growls, which are produced by modifying the length and shape of the vocal tract \cite{nieto2013}. These screamed vocals serve one of two purposes: they are either low and beast-like to accentuate the aggressive, darker themes of heavy metal, or high and screechy, to stand out from the otherwise aggressive sounds of the distorted electric guitar \cite{deathmetalmusic}. In this paper we explore methods to detect and classify the type of vocal technique being used by a vocalist.

The automatic identification of different type of vocal techniques in heavy metal could, for instance, inform genre classification systems and aid music recommendation systems based on preference for a specific vocal type. Vocal detection for heavy metal music could also improve vocal extraction as well as (lyrics) transcription for this genre.

\begin{figure*}[t]
\centering
\includegraphics{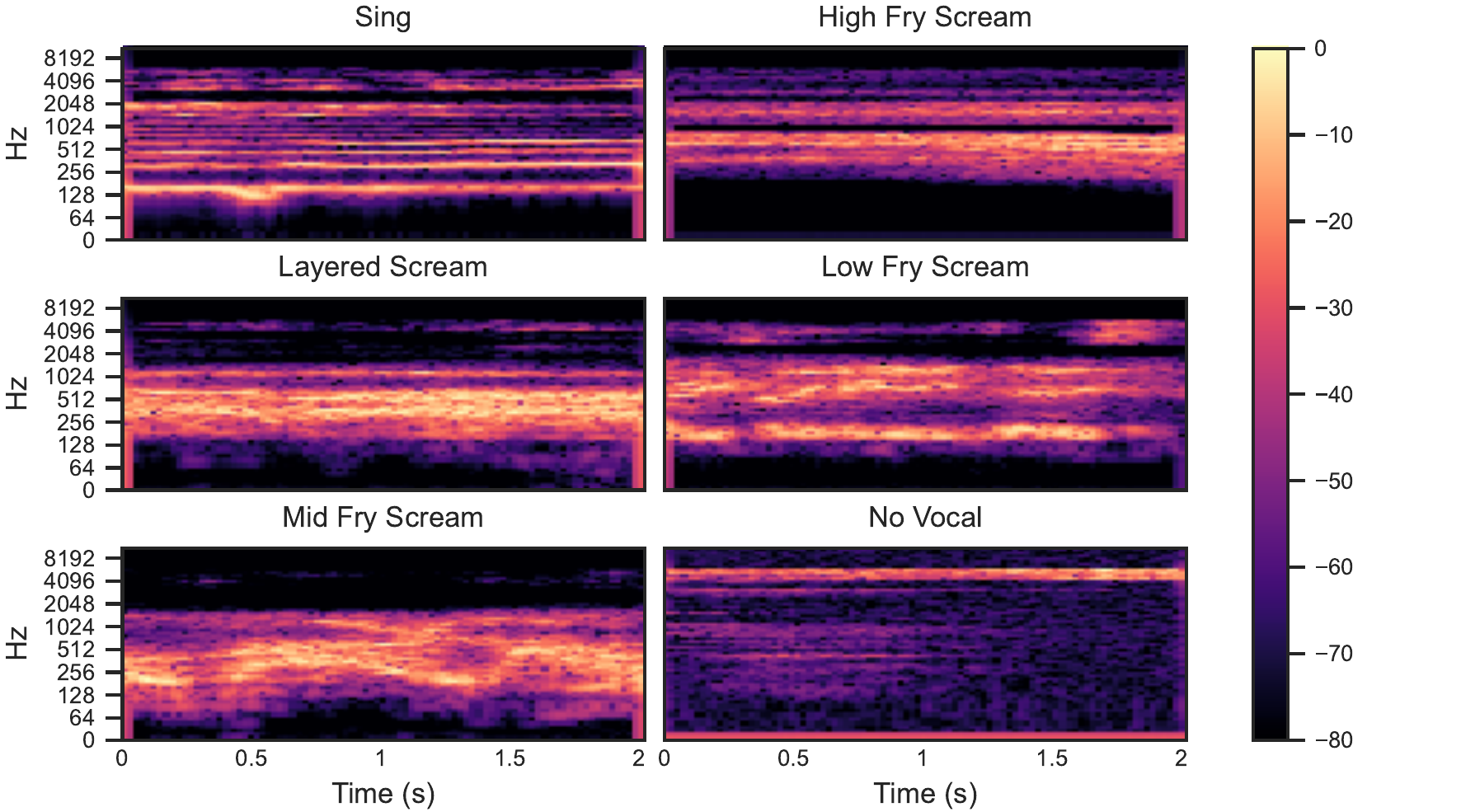}
\caption{Example spectrogram representation of different screams.\label{fig:classwise-spectrogram}}
\end{figure*}

Nieto introduced the term `Extreme Vocal Effects' or EVEs to describe the vocal styles present in heavy metal\cite{nieto2008}. These EVEs fall into 3 main categories:
\begin{compactitem}
  \item \textit{Growls}: Growls are common in death metal. They are very noisy and the fundamental frequency is rarely perceived. They are usually loud and produce a high amount of spectral variation \cite{smialek2012formants,nieto2013} 
  \item \textit{Fry Screams}: Fry screams are similar to growls, but are brighter and not as loud. They are produced by a series of irregularly spaced glottal pulses that are induced by inhaling or exhaling \cite{ishi2007}
  \item \textit{Rough Vocals}: Rough vocals are obtained by adding variations in the vocal tract to obtain a harmonically richer spectrum \cite{nieto2013,smialek2012spectrographic}. This is much more common in rock than in metal (e.g., for bands such as \textit{Foo Fighters} and \textit{Breaking Benjamin}).
\end{compactitem}
 {Figure~\ref{fig:classwise-spectrogram} shows a sample spectrogram for each class. Distinct patterns in the low and mid fry scream can be observed that distinguish them from the other types of screams. The high screams occupy a higher portion of the spectrum as well. It is important to note that, in these examples, the mid fry scream appears to have lower frequency content than the low fry scream. This is because these are examples chosen from different vocalists, and the perceived type of scream varies according to factors discussed in Sect.~\ref{sec:dataset}.}

Some subgenres of metal also involve sung or `clean' vocals. In this paper, `screams' and `growls' will be used to describe the overall style of distorted heavy metal vocals, and `clean' will be used to describe sung vocals. The term growl usually refers to the low pitched, rough sounds uttered by animals. Humans occasionally use growl-like voices to express strong emotions. Examples of `growl' phonations have been seen across the genres of jazz, blues, gospel, samba, country and pop. In ethnic music, the growl is found in \emph{umngqokolo} (the vocal tradition of the Xhosa people), and throat singing (Tuvan and Mongolian)\cite{sakakibara2004}. However, in recent times growls are most strongly associated with metal vocals.

Extreme metal screams can be performed by either inhaling or exhaling which has a noticeable effect on the timbre of the sounds produced. However, in most modern metal, screams are produced by exhaling, and so our work will focus on these types of screams.

The remainder of this paper is structured as follows. After an overview of related work in Sect.~\ref{sec:relatedwork}, a new publicly available dataset is introduced in Sect.~\ref{sec:dataset}. We describe several benchmark systems for detection and classification in Sect.~\ref{sec:method} and present the corresponding results in Sect.~\ref{sec:results}. The conclusion in Sect.~\ref{sec:conclusion} summarizes the main contributions in gives a brief outlook of future work.

\section{Related Work}\label{sec:relatedwork}
While there exists, to the best knowledge of the authors, no previous work on the automatic categorization of  heavy metal vocals, one related field is the detection of screams in urban environments.

In the previous section, we introduced different types of screams in metal music. Here, we will discuss past work with scream detection in general, followed by related work on screamed vocals in heavy metal music.

Prior work in detecting screams aims at the detection and localization of screams in urban sound, the detection of screams in subways, scream and shout recognition in noise, and scream detection for home applications. Various approaches were taken to achieve these tasks. Huang et al.\ used  Mel Frequency Cepstral Coefficients (MFCCs) and Support Vector Machine (SVM) to classify blocks of audio captured from a microphone array into the two classes scream and non-scream \cite{huang2010}. Rabaoui et al.\ use a one class SVM to classify a sound into 9 categories, including scream, gunshot, explosion, door slam or dog barks,
using features such as spectral centroid, spectral roll-off, zero-crossing rate, MFCCs and Linear Predictive Coding Coefficients (LPCCs) \cite{rabaoui2007}. They also included the first and second derivatives of these features, but determined that they were not helpful in improving performance. Lafitte et al.\ used a deep neural network approach with MFCCs to detect shouted voice/screams in subway trains, and classify audio into shout, conversation and noise
\cite{lafitte2016}. 
Other work in detecting screams in noise also uses MFCC and spectral entropy features with GMM classifiers to achieve this task \cite{nandwana2015,pohjalainen2011,hayasaka2017}. The best performing of these methods was able to achieve equal error rates (EERs) of 0.3\% and 0.8\% under 0dB and -5dB signal to noise ratio (SNR) conditions. This approach, while useful in identifying screams in noisy conditions, cannot be translated well to detecting screams in music since the noise added was that of subway stations, trains and air conditioners.

Most work related to heavy metal vocals focuses on the physiology of screamed vocals \cite{eckers2009,sakakibara2004,ribaldini2019,loscos2004}, their spectral properties \cite{guzman2019}, and exploratory acoustic feature analyses \cite{kato2013,sakakibara2004}. There has been limited work on detecting and classifying the types of vocals present in heavy metal. Nieto uses k-means clustering to group different vocal styles into the three classes \textit{Growl}, \textit{Fry Scream}, and \textit{Roughness} \cite{nieto2013}. The dataset used consisted of labeled recordings of the 6 vocalists' screams. While this work was successful at grouping similar classes together, it could not predict the type of EVE present. 
Due to a lack of data with start and end times of vocal events annotated, a sliding window approach similar to Huang, where the scream detection algorithm is applied to every block in a sliding window to determine the start and end times of a scream \cite{huang2010} could not be implemented, and hence identifying when a scream occurs, or identifying what different kinds of screams are present within one file were not possible.



\section{Dataset}\label{sec:dataset}

Currently, there exists no publicly available dataset with annotated vocals for heavy metal. To enable this study, as well as to facilitate future research on this topic, we present the newly created \emph{Metal Vocal Dataset} (MVD). This dataset consists of 57 songs from 34 bands and 47 albums. The list of songs can be found in the appendix. Most of these songs were released during the last two decades, since use of vocal effects beyond Mid Fry screams has increased in this period.

A playlist containing all the songs present in the dataset was created.\footnote{\href{https://www.youtube.com/playlist?list=PLnkRJFUtBDzWOEnVOiWTVxGOWD70LDwtC}{https://tinyurl.com/metal-vocal-dataset-playlist}} The distribution of the songs selected for the dataset based on the year of release is shown in Figure~\ref{fig:release-year-bar}.

The annotations have been released under the MIT license and are available online.\footnote{\href{https://github.com/VedantKalbag/metal-vocal-dataset}{https://github.com/VedantKalbag/metal-vocal-dataset}} The audio files themselves are not included, but can be retrieved using a script provided in the repository.

\subsection{Data Selection}
The songs selected were from genres such as death metal, groove metal, progressive metal, black metal, and metal core. The traditional subgenres of death metal, black metal and groove metal were included as they contain mostly one class of screams (mid fry screams), while modern subgenres such as metal core and progressive metal were chosen since a wide variety of vocal effects are used in these genres. The songs were selected with the aim to capture a wide variety in vocal styles and are listed in a playlist.\footnotemark[1]

\subsection{Dataset Statistics}
\begin{figure}
\centering
\includegraphics[width=\columnwidth,angle=0]{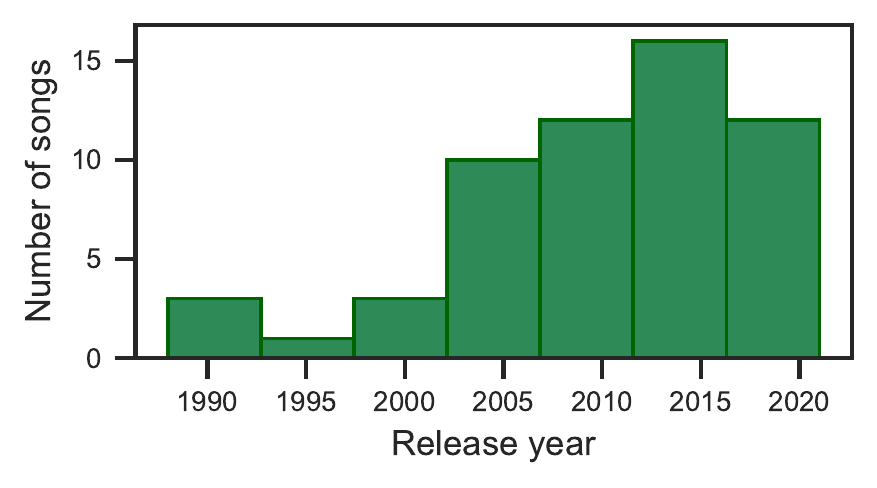}
\caption{Distribution of dataset based on release year.\label{fig:release-year-bar}}
\end{figure}

The distribution of the songs selected for the dataset based on the year of release is shown in Fig.~\ref{fig:release-year-bar}. The increase for more recent years reflects the increased use of vocal effects beyond mid fry screams.

\begin{figure}
\centering
\includegraphics[width=\columnwidth,angle=0]{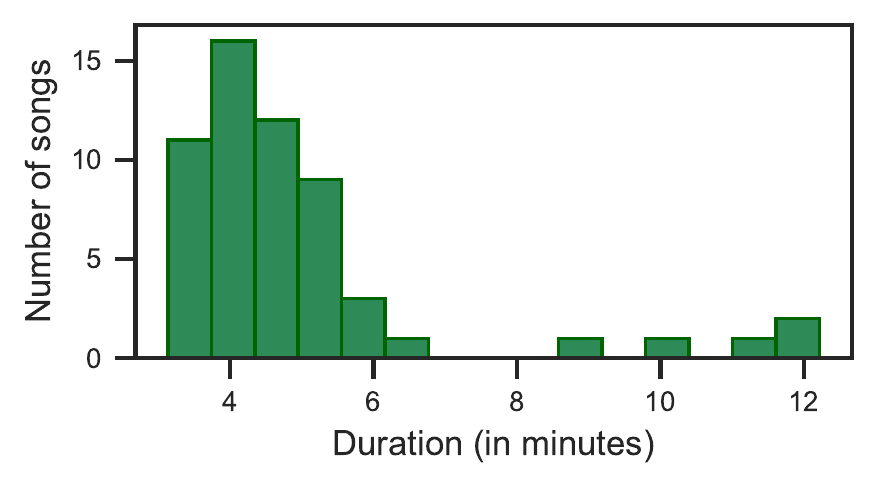}
\caption{Distribution of dataset based on song length in minutes.\label{fig:duration-bar}}
\end{figure}

\begin{figure}[t]
\centering
\includegraphics[width=\columnwidth,angle=0]{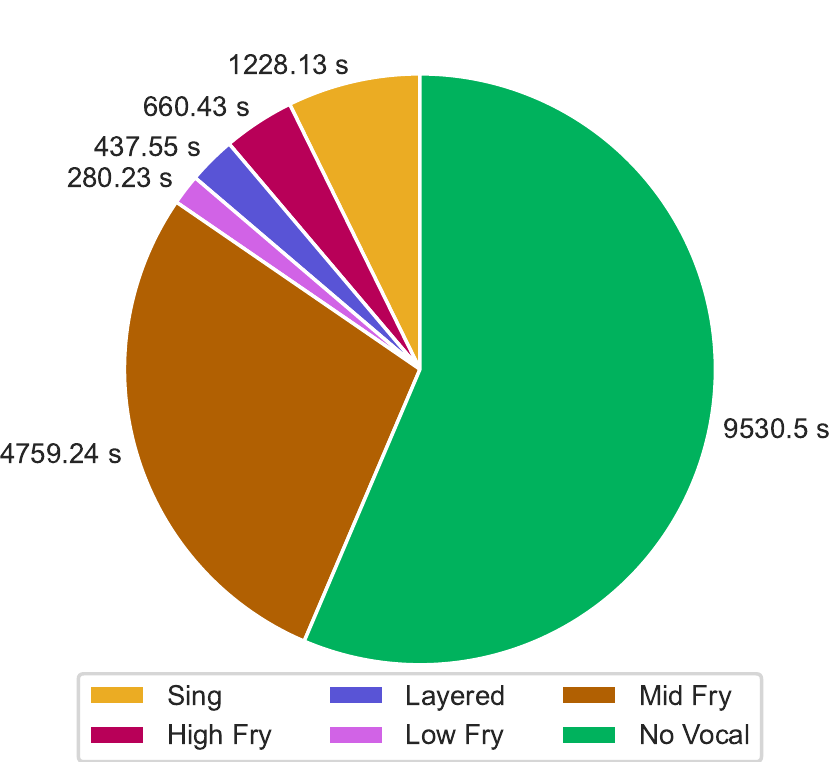}
\caption{Distribution of dataset based on total time per class.\label{fig:duration-pie}}
\end{figure}

There are a total of \unit[281.6]{min} of audio across the 6 classes (including the `no vocal' class). The class distribution in the dataset is visualized in Fig.~\ref{fig:duration-pie} based on the total time annotated in seconds. The Mid Fry scream is the largest part of the dataset; although the songs were selected carefully to contain all different classes, the Mid Fry scream is most prevalent in modern metal music.

\subsection{Data Split}
The data was split into 3 subsets for training, testing, and validation. This was done after division of audio files into 1 second blocks as described further in Sec.~\ref{sec:method}. Since the class distribution was heavily skewed towards blocks labeled `no vocals', the dataset was undersampled to balance out classes. All classes that had more samples than the class with minimum samples were undersampled to the nearest thousand, for both the 3-class as well as the 6-class problem.

The data is accompanied by a recommended split into the subsets train, validation, and test (approx.\ 70:15:15). The data was split such that no band's songs are present in both the training and test/validation sets. Undersampling was applied before the split to balance the class distribution, as undersampling after the split would lead to considerably smaller test and validation sets. The blocks were first divided into an approx.\ 70:30 split, ensuring that no band was present in both subsets. This split at a band level was done to avoid overfitting any one vocalist/band and hence giving false results. The 30\% split was then divided into two equal subsets at random. This was done because when restricting one band to be in either the test or validation set only drastically reduced the size of these sets, and would render them useless.
In addition, a recommended split with imbalanced class distribution containing all data is provided as well. 

\begin{figure*}
\centering
\includegraphics[width=5in,angle=0]{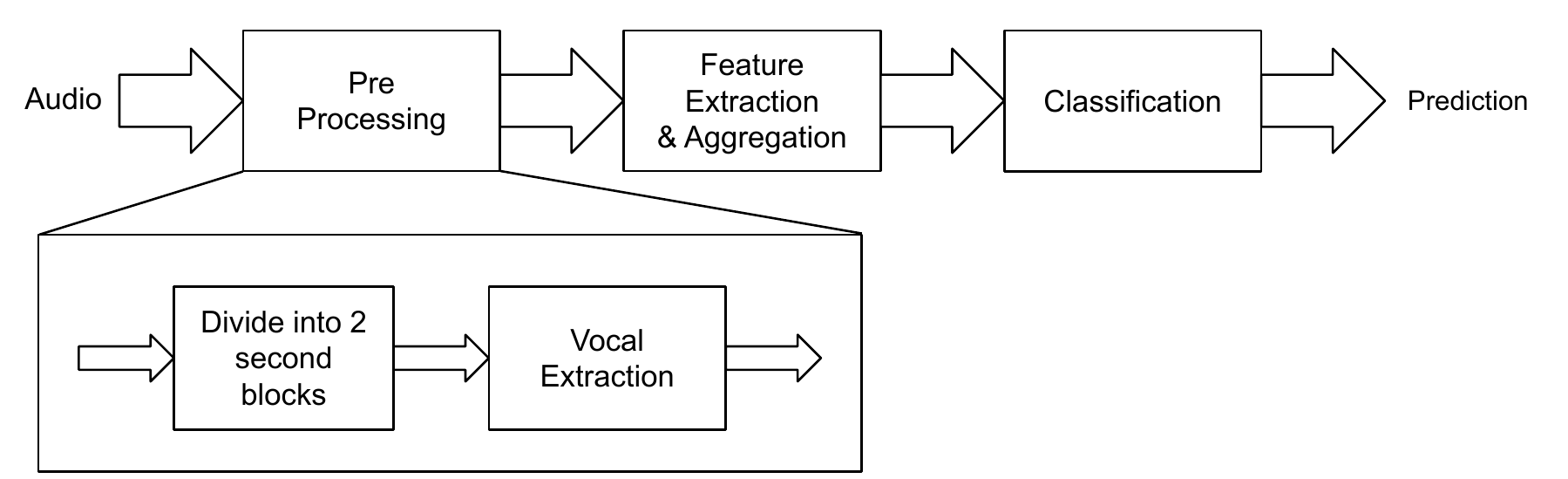}
\caption{{Block diagram of benchmark systems.}\label{fig:block-diagram}}
\end{figure*}
\subsection{Annotation Methodology}\label{subsec:annotation-methodology}

Since most screams in modern metal are variations of a fry scream, we have focused on these for our dataset. The variations are caused by a change in the shape and length of the vocal tract, where lengthening the vocal tract makes the scream sound lower, and vice versa. We have defined 3 fry scream categories based on the perceived sound: High, Mid, and Low. 
Thus, the vocal events were annotated with the following class labels: 
\textit{Sing}, \textit{High Fry} scream, \textit{Mid Fry} scream, \textit{Low Fry} scream, and \textit{Layered} scream. The class labeled `layered' contains combinations of 2 or more other classes simultaneously (e.g., Mid Fry screams and Sing, or both High and Low Fry screams).

These songs were manually annotated using \textit{Sonic Visualiser} so that the maximum time difference between the start or end of a vocal event and the annotation is less than \unit[0.5]{s}. The start and end points of the vocal event were localized visually based on the spectrogram of the audio file and validated aurally.

An important consideration is that the categorization of some screams is subjective, and two individuals may assign class labels differently. For example, a `low-sounding' Mid Fry scream could be perceived as a `high-sounding' Low Fry scream, and vice versa. 
As the main criteria for labeling the screams, the vowel characteristics of the sound were used. Typically, a Low Fry scream will have dark vowel characteristics (\textit{/o/} or \textit{/u/}), a Mid Fry scream will have vowel characteristics similar to \textit{/a/}, and a High Fry scream will have characteristics around \textit{/e/} or \textit{/i/}.
The labels were, thus, assigned based on how the scream sounded with respect to the perceived vowel characteristics;  for instance, a scream with prominent low frequencies and vowel characteristics of \textit{/u/} or \textit{/o/} was labeled as a Low Fry scream.

\section{Benchmark Methods}\label{sec:method}
A block diagram of the systems created as a benchmark for future work is shown in Fig.~\ref{fig:block-diagram}, and is described in detail in the following.

\subsection{Pre-processing}\label{subsec:pre-processing}
The audio files were passed through the Spleeter source separation algorithm \cite{spleeter2020} to separate the vocals from the other components and then divided into overlapping blocks of length \unit[2]{s} with a \unit[1]{s} hop size. Each \unit[2]{s} block is one observation to be classified. All audio files were resampled to a sample rate of \unit[44100]{Hz}, normalized and downmixed to mono.

\subsection{Input Representation}\label{subsec:feature-extraction}
The baseline set of features consists of low level temporal and spectral features that are commonplace in Music Information Retrieval tasks. These features are: 13 MFCCs and Delta MFCCs, RMS, ZCR, Spectral Centroid, Contrast, Flatness and Roll-off (for a feature definition see \cite{lerch_introduction_2012}). These features were extracted using the Librosa python library \cite{mcfee2015librosa}, with a window size of \unit[2048]{samples} and a hop size of \unit[1024]{samples}. 
In addition, VGGish features \cite{hershey2017vgg} and the Log-Mel Spectrogram were extracted.

We divide these features into the following feature sets:
\begin{compactenum}
    \item Feature Set 1: 13 MFCCs, Delta MFCCs, RMS, ZCR, Spectral Centroid, Contrast, Flatness and Roll-off
    \item Feature Set 2: VGGish Features
    \item Feature Set 3: 13 MFCCs and Delta MFCCs only
    \item Feature Set 4: RMS, ZCR, Spectral Centroid, Contrast, Flatness and Roll-off
    \item Feature Set 5: Log Mel Spectrogram
\end{compactenum}


\subsection{Feature Aggregation}\label{subsubsec:baseline-features}
All features in Feature Set 1 were aggregated by taking the mean and standard deviation across each audio block (with duration \unit[2]{s}). The features in Feature Set 1, 2, 3, and 4 were all z-score normalized across the entire training set to return a feature vector with 0 mean and unit standard deviation. The mel spectrogram input was converted to log scale before use.

\subsection{Classifiers}\label{subsec:classifiers}
Two multi-class classifiers were used to classify each audio block based on the feature vector. 
The different classifiers used are a Support Vector Machine (SVM) and a Convolutional Neural Network (CNN). {The CNN consists of 3 convolutional layers with dimensions 256, 512, and 1024, each followed by max pooling, respectively, 3 dense layers with dimensions 256, 64, and 16, and an output layer.}

\section{Experiments}\label{sec:evaluation}

The system was tested for two different sets of labels: a 3 class problem (sing, scream, no vocal), as well as a 6 class problem (containing all the 5 labels from the dataset as well as no vocal).


\subsection{Experiment 1: 3-Class Problem}
All scream classes are combined into a single class, resulting in the target set of classes \textit{Sing}, \textit{Scream}, and \textit{No Vocal}. The following configuration were evaluated:
\begin{compactenum}
  \item Feature Set 1 + SVM
  \item Feature Set 2 + SVM
  \item Feature Set 3 + SVM
  \item Feature Set 4 + SVM
  \item Feature Set 5 + CNN
\end{compactenum}

\subsection{Experiment 2: 6-Class Problem}
As opposed to Experiment 1, Experiment 2 treats each scream class separately, resulting in the target set of classes \textit{Sing}, \textit{Low Fry}, \textit{Mid Fry}, \textit{High Fry}, \textit{Layered}, and \textit{No Vocal}.  
This experiment investigates the two best-performing SVM configurations {and the CNN configuration from Exp.~1:}
\begin{compactenum}
  \item Feature Set 1 + SVM
  \item Feature Set 2 + SVM
  {\item Feature Set 5 + CNN}
\end{compactenum}

\subsection{Performance Metrics}\label{subsec:metrics}
The performance metrics used in this study are:
\begin{compactenum}
  \item Accuracy: \textit{acc}
  \item Macro-Accuracy: \textit{bal-acc}
  \item Balanced F1 Score: \textit{f1}
\end{compactenum}
These metrics were computed with the sklearn python library \cite{scikit-learn}.

\section{Benchmark Results}\label{sec:results}

The results of both the 3-class and 6-class classification problem are presented below, followed by a discussion of the results. The results for a 3 class implementation, with blocks being classified into sing, scream and no vocal are compared to a 6 class implementation, where the audio block was classified into Sing, Low Fry scream, Mid Fry scream, High Fry scream, Layered screams and No Vocal.

\subsection{Experiment 1: 3-Class Results}\label{subsec:feature-results}
\begin{figure}
\centering
\includegraphics[width=\columnwidth,angle=0]{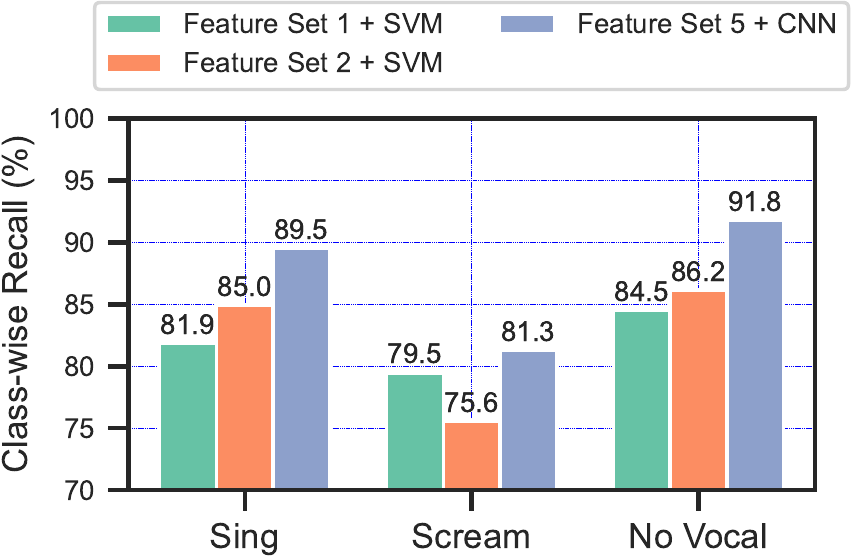}
\caption{Class-wise recall for the 3-class problem.\label{fig:classwise-accuracies}}
\end{figure}

The results for each experiment are shown in Table~\ref{tab:3class-results-features} and the class-wise recall of each combination are shown in Fig.~\ref{fig:classwise-accuracies}.

\begin{table}
 \begin{tabular*}{\columnwidth}{l|  @{\extracolsep{\fill}}c|c|c}	
  \textbf{Configuration} & \textbf{acc} & \textbf{bal-acc} & \textbf{f1}\\
  \hline
  \hline
  Feature Set 1 + SVM & 82.20 & 82.10 & 82.18\\
  \hline
  Feature Set 2 + SVM & 82.06 & 82.23 & 82.10\\
  \hline
  Feature Set 3 + SVM & 77.12 & 76.95 & 77.21\\
  \hline
  Feature Set 4 + SVM & 79.55 & 79.40 & 79.60\\
  \hline
  {Feature Set 5 + CNN} & \bf{87.33} & \bf{87.58} & \bf{87.42}\\
 \end{tabular*}
 \caption{Results for the 3-class problem in Exp.~1 (values shown in \%)}
 \label{tab:3class-results-features}
\end{table}

\begin{figure}
\centering
\includegraphics[width=\columnwidth,angle=0]{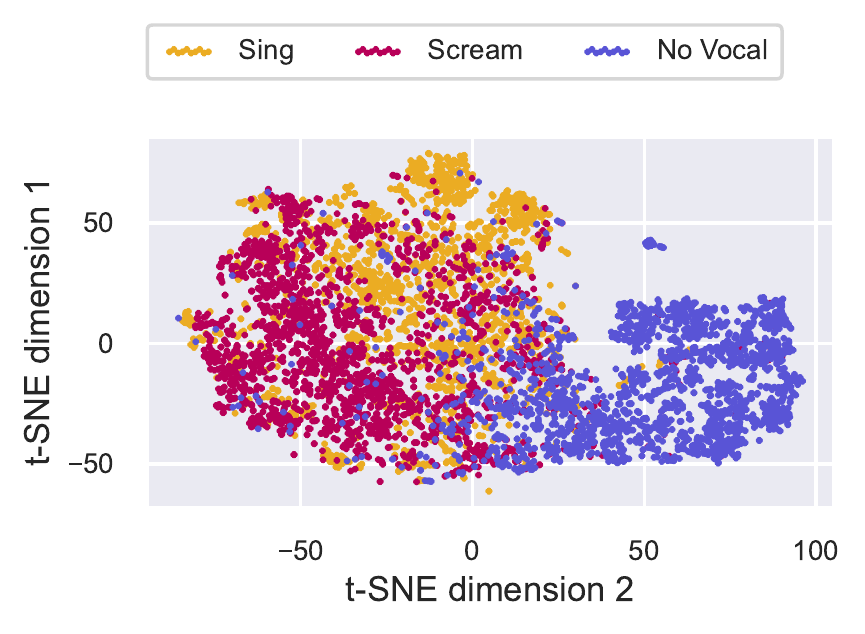}
\caption{t-SNE projections of the feature space (Feature Set 1).\label{fig:tsne}}
\end{figure}

{Figure \ref{fig:tsne} shows the t-SNE plot of Feature Set 1, and we can see a distinction between the 3 different classes, although some overlap between the classes \textit{Sing} and \textit{Scream}.}
We can make the following observations. First, combined Feature Set 1 outperforms Feature Sets 3 and 4 with a gap of roughly 5\%. This is expected as these sets are subsets of Feature Set 1. 
Second, the combined Feature Set 1 and the VGGish Feature Set 2 show the best performance and perform similarly with recall above 82\%. This means that the VGGish features, trained on a different task, contain a similar, semantically meaningful, information for classification as the combination of common baseline features. To a degree it is surprising that Feature Set 2 does not clearly outperform the traditional feature set as VGGish features have been shown to be powerful in music tasks such as musical instrument classification \cite{gururani_attention_2019}.
Third, the results show that the CNN with spectrogram input is able to detect the presence of screams with 87.6\% balanced accuracy, which is notably higher accuracy than any SVM-based approach. It seems that the CNN is able to utilize the information in the spectrogram and is able to detect spectral patterns efficiently. 

\subsection{Experiment 2: 6-Class Results}\label{subsec:vggish-results}

The results of the 6-class problem are given in Table~\ref{tab:6class-results-features}. We can observe that the performance is considerably lower for the 6-class problem with the two top-performing feature sets from Exp.~1. The VGGish features in Feature Set 2 seem to slightly outperform the low-level Feature Set 1. {The CNN did not perform as well as the combination of VGGish features and SVM; the results of the CNN appear to be biased towards High Fry screams (see below).}
\begin{table}
 \begin{tabular*}{\columnwidth}{l|  @{\extracolsep{\fill}}c|c|c}	
  \textbf{Configuration} & \textbf{acc} & \textbf{bal-acc} & \textbf{f1}\\
  \hline
  \hline
  Feature Set 1 + SVM & 44.24 & 41.92 & 38.03\\
  \hline
  {Feature Set 2 + SVM} & \bf{45.53} & \bf{45.91} & \bf{40.13}\\
  \hline
  {Feature Set 5 + CNN} & 42.89 & 40.87 & 38.79
 \end{tabular*}
 \caption{Results for the 6-class problem in Exp.~2 (values shown in \%)}
 \label{tab:6class-results-features}
\end{table}

\begin{figure}[t]
\centering
\includegraphics[width=.99\columnwidth,angle=0]{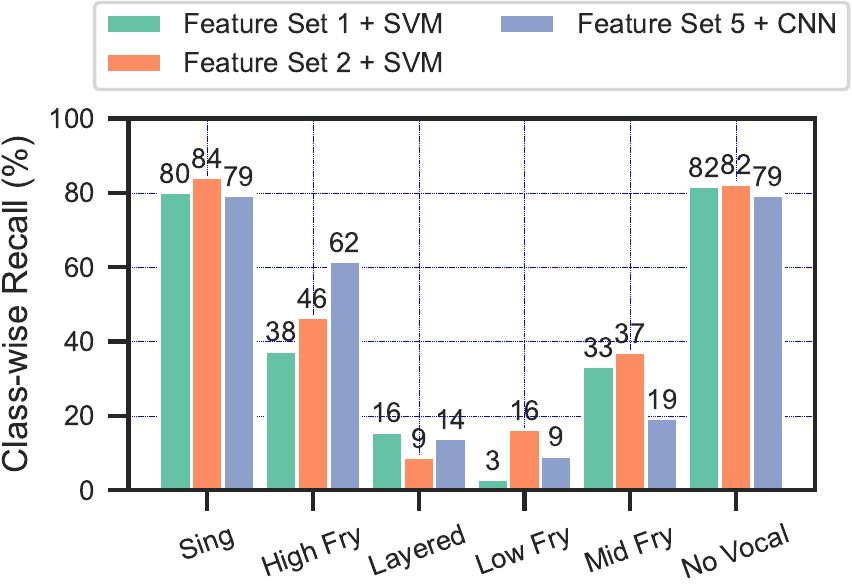}
\caption{Class-wise recall for the 6-class problem.\label{fig:6classrecall}}
\end{figure}

Looking at the class-wise recall in Fig.~\ref{fig:6classrecall}, we observe that the systems could still identify the sung vocal and absence of vocals with high accuracy in the same range as the 3 class results shown above, however, they could not accurately distinguish between the different types of screams. {We also see that the recall of the High Fry scream in the CNN is significantly higher than the other experiments, which is due to the classifier predicting most screams to be High Fry screams.}
%
 %

\begin{figure}[t]
\centering
\includegraphics[width=.99\columnwidth,angle=0]{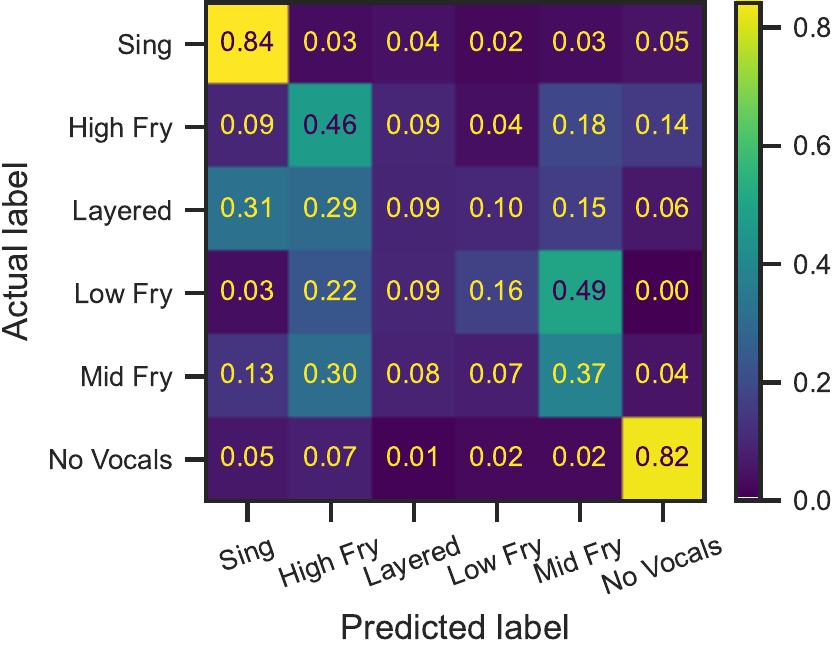}
\caption{Confusion matrix for the 6-class problem (SVM).\label{fig:cm-6class}}
\end{figure}

Investigating the confusion matrix in Fig.~\ref{fig:cm-6class} gives us more details of the problem with the screams. We can see that several classes are being predicted incorrectly. Blocks labeled `Layered' were often predicted as other classes, especially 'Sing' and 'High Fry' this could be because the layered class contains combinations of different classes, including the `Sing' vocals. We also see that `Low Fry' screams are often predicted as `Mid Fry' due to the high degree of overlap between these classes in the feature space.

\section{Conclusion}\label{sec:conclusion}
We introduced a new annotated dataset to aid and encourage further research in vocal detection in heavy metal music. Both the dataset and code have been made publicly available. While targeting scream detection, the dataset is also suitable for related tasks such as Vocal Activity Detection.

We presented a set of benchmark experiments on the automatic detection and classification of vocals in heavy metal music with the presented dataset. In these experiments, various temporal, spectral, and cepstral, and VGGIsh features were evaluated and compared with a CNN with log-mel spectrogram input. 

In conclusion, with the dataset presented in this paper, we were able to detect the presence of vocal events and classify them into sung vocal and screamed vocal with good accuracy. However, the same cannot be said for classifying the screams into the different types, as the different scream classes overlap within the feature space and cannot be separated easily. Thus, the dataset provides a new challenging task that can currently not be solved with satisfying results.

\subsection{Future Work}\label{subsec:futurework}
There is anecdotal evidence online and within the heavy metal community for additional categories of vocal effects such as `guttural vocals' and `pig squeals'. Pending further investigation into this, we plan the extension of the dataset with additional audio files as well as extending the annotations to include these additional subsets of extreme vocal effects.  

At present, the dataset has limited samples containing clean vocals sung over distorted instrumental sections, as most of the sections containing clean vocals in the songs used were also softer in nature. The dataset also has fewer samples of Low and High Fry screams (this is representative of their use in modern metal), and can be expanded upon by including further examples of these vocals.


\bibliography{sections/bibliography}

\clearpage
\section{Appendix}\label{sec:appendix}
The following songs were included in the dataset (Song No.~Artist~--~Song Name):\\
 
\begin{compactenum}
\item Abbath -- Ashes Of The Damned
\item After The Burial -- Lost In The Static
\item Amon Amarth -- Destroyer of the Universe
\item Amon Amarth -- Live For The Kill
\item Amon Amarth -- Twilight Of The Thunder God
\item Be'lakor -- Venator
\item Behemoth -- Ecclesia Diabolica Catholica
\item Behemoth -- Bartzabel
\item Behemoth -- Blow Your Trumpets Gabriel
\item Born of Osiris -- White Nile
\item Cannibal Corpse -- High Velocity Impact Spatter
\item Children of Bodom -- Under Grass And Clover
\item Children of Bodom -- Living Dead Beat
\item Children Of Bodom -- Are You Dead Yet
\item Children of Bodom -- Sixpounder
\item Children Of Bodom -- Everytime I Die
\item Children Of Bodom -- In Your Face
\item Dark Tranquillity -- Lost to Apathy
\item Dark Tranquillity -- Atoma
\item Death -- Pull the Plug
\item Death -- The Philosopher
\item Decapitated -- Kill The Cult
\item Decapitated -- Blood Mantra
\item Ensiferum -- In My Sword I Trust
\item Enslaved -- Caravans To The Outer Worlds
\item Godless -- Deathcult
\item Gojira -- Stranded
\item Gojira -- Silvera
\item Immortal -- Northern Chaos Gods
\item In Flames -- Cloud Connected
\item Lamb of God -- Memento Mori
\item Lamb of God -- Laid to Rest
\item Lamb of God -- Omerta
\item Lamb of God -- Now You've Got Something to Die For
\item Lamb of God -- The Faded Line
\item Ne Obliviscaris -- Pyrrhic
\item Ne Obliviscaris -- And Plague Flowers the Kaleidoscope
\item Nevermore -- Born
\item Of Mice \& Men -- Bones Exposed
\item Of Mice \& Men -- Obsolete
\item Opeth -- Blackwater Park
\item Parkway Drive -- Carrion
\item Rings of Saturn -- Senseless Massacre
\item Slayer -- War Ensemble
\item Slayer -- South Of Heaven
\item Slipknot -- Psychosocial
\item Suffocation -- Clarity Through Deprivation
\item Suicide Silence -- No Pity for a Coward
\item Suicide Silence -- Disengage
\item Suicide Silence -- You Only Live Once
\item Suicide Silence -- Slaves To Substance
\item Tesseract -- Nocturne
\item Textures -- Storm Warning
\item Textures -- Old Days Born Anew
\item Thy Art Is Murder -- Reign Of Darkness
\item Veil of Maya -- Overthrow
\item Wintersun -- Time
\end{compactenum}
\end{document}